\documentclass[%
 reprint,
 amsmath,amssymb,
 aps,
]{revtex4-2}

\usepackage{graphicx}
\usepackage{dcolumn}
\usepackage{bm}

\usepackage{comment}

\begin{document}

\preprint{APS/123-QED}

\title{Deep Learning in Classical X-ray Ghost Imaging for
Dose Reduction}

\author{Yiyue Huang}
\email{yiyue.huang@anu.edu.au}
\altaffiliation{Dept. of Materials Physics, RSPhys, The Australian National University, Canberra 2600 ACT, Australia}

\author{Philipp D. Lösel}%
\affiliation{Dept. of Materials Physics, RSPhys, The Australian National University, Canberra 2600 ACT, Australia}


\author{David M. Paganin}
\affiliation{School of Physics and Astronomy, Monash University, Clayton 3800 VIC, Australia}%

\author{Andrew M. Kingston}
\email{andrew.kingston@anu.edu.au}
\affiliation{Dept. of Materials Physics, RSPhys, The Australian National University, Canberra 2600 ACT, Australia }

\thanks{PDL and AMK also at: National Laboratory for X-ray Micro Computed Tomography, Advanced Imaging Precinct, The Australian National University, Canberra 2600 ACT, Australia.}%


\date{\today}

\begin{abstract}
Ghost imaging (GI) is an unconventional technique that combines information from two correlated patterned light fields to compute an image of the object of interest. A standard pixelated camera records the structure of one light field (that does not interact with the object), and a bucket detector (or single-pixel camera) measures the total intensity of the second light field that is transmitted or scattered by the object. GI can be performed with visible light as well as penetrating radiation such as x-rays, electrons, etc. Penetrating radiation is usually ionizing and damages biological specimens; therefore, minimising the dose of this radiation in a medical or biological imaging context is important. GI has been proposed as a potential way to achieve this. With prior knowledge of the object of interest, such as sparsity in a specific basis (e.g., Fourier basis) or access to a large dataset for neural network training, it is possible to reconstruct an image of the object with a limited number of measurements. However, low sampling does not inherently equate to low dose. Here, we specifically explore the scenario where reduced sampling corresponds to low-dose conditions. In this simulation-based paper, we examine how deep learning (DL) techniques could reduce dose in classical x-ray GI. Since GI is based on illumination patterns, we start by exploring optimal sets of patterns that allow us to reconstruct the image with the fewest measurements, or lowest sampling rate, possible. We then propose a DL neural network that can directly reconstruct images from GI measurements even when the sampling rate is extremely low. We demonstrate that our deep learning-based GI (DLGI) approach has potential in image reconstruction, with results comparable to direct imaging (DI) at the same dose. However, given the same prior knowledge and detector quantum efficiency, it is very challenging for DLGI to outperform DI under low-dose conditions. We discuss how it may be achievable due to the higher sensitivity of bucket detectors over pixel detectors.

\end{abstract}

\maketitle


\section{\label{sec:Introduction}Introduction}




X-ray technology is widely used to produce 2D and 3D images of structures inside the human body. Nevertheless, since x-rays are a form of ionizing radiation, overdoses of medical x-ray imaging may result in higher cancer incidence rates \cite{linet2012cancer}. Consequently, there must be a trade-off between gaining more information or achieving higher image quality and minimising x-ray dose. In excised samples, an excessive dose can damage the structure, thus distorting the resulting images. So, “how can we reduce the dose while still obtaining a useful image?” becomes an important question. It has been proposed that ghost imaging has the potential to achieve this \cite{kingston2021inherent}.

Ghost imaging (GI) is a new paradigm in imaging that was first realised using entangled photon pairs in the field of quantum optics late last century (see \cite{padgett2017introduction}). Classical variants were later realised using thermal light with structured/patterned illumination since only the position correlation property of the entangled photons was required \cite{padgett2017introduction}. The GI measurement process is outlined in Sec. \ref{sec:bg-compgi} and the image computation is detailed in Sec. \ref{sec:CGImethod}. Classical GI has now been extended from visible light to penetrating radiation such as x-rays \cite{yu2016fourier,pelliccia2016experimental}, electrons \cite{li2018electron}, neutrons \cite{kingston2020neutron}, etc. The x-ray ghost imaging technology has been utilised in two-dimensions by Zhang et al. \cite{zhang2018tabletop} and Pelliccia et al. \cite{pelliccia2016experimental}, as well as in three-dimensions by Kingston et al. \cite{kingston2018ghost}. Without prior knowledge of the specimen, it is challenging to recover high quality images using classical GI. Kingston et al. showed that direct imaging or a scanning probe will always produce better results than classical GI under the same photon-shot noise \cite{kingston2021inherent}. However, classical GI can become advantageous when we have prior knowledge of the imaging object. It becomes theoretically possible to reconstruct the object with fewer measurements, thus resulting in reduced radiation dose, preventing the object from being damaged in biological-imaging settings. One way to achieve this is through compressed sensing (CS), in which an image is assumed to be sparse in a particular transform space or domain, (e.g., the Fourier space), allowing images to be reconstructed under a low sampling rate \cite{candes2008introduction}. However, this requires an extremely sparse object image in some particular basis, and this is rarely satisfied. Another method is through artificial intelligence techniques such as deep learning \cite{alzubaidi2021review}, using a large dataset that encapsulates the prior knowledge to train a neural network. For example, Wang et al. have developed a neural network with a structure based on eHoloNet \cite{wang2018eholonet} that outperforms other ghost imaging reconstruction methods even when the sampling rate is low \cite{wang2019learning}.

The purpose of this paper is to explore the possibility of dose reduction in x-ray ghost imaging through deep learning. There have been many studies investigating GI with deep learning (e.g. \cite{wang2019learning,liu2021computational,lyu2017deep}), even in the context of x-ray imaging \cite{li2023nanoscale} and neutron imaging \cite{he2021single}. Lyu et al. \cite{lyu2017deep} showed that their deep learning ghost imaging approach outperforms classical compressive sensing techniques at extremely low sampling rates. Inspired by this, we aim to further explore its application in low-dose conditions. While a low sampling rate is often associated with a low-dose condition, it is not a direct correlation; it is possible to have a low sampling rate but a high dose, i.e., a significant amount of x-ray radiation is used. Hence, this work explores the application of deep learning ghost imaging (DLGI) in extremely low-dose conditions. We first assume that fewer measurements lead to a lower dose under the same conditions. We then attempt to find optimal illumination patterns and develop a neural network that minimises the number of measurements while still obtaining sufficient image information with a reasonable computation time. Following that, we compare our DLGI recovered images with images produced by conventional direct imaging subjected to the same total dose, to see if dose reduction can be achieved through this DLGI approach to classical x-ray ghost imaging. Note that in our simulations, we assume the same detector quantum efficiency in the comparison, where DLGI will be at an inherent disadvantage compared to direct imaging. In practice, we would anticipate an increase in measured flux in the GI data while maintaining the same dose on the object (see Sec. \ref{sec:5C}).

\section{\label{sec:Background}Background}
In this section, we provide relevant background information. We begin by introducing the principle of classical ghost imaging and computational ghost imaging. We then look at how deep learning works and how ghost imaging may benefit from it. Following that, we introduce the noise model for x-ray measurements that is needed to enable low-dose simulations. Further, we outline principal component analysis, which can be used to construct a set of potential optimal illumination patterns for GI.

\subsection{Computational Ghost Imaging}
\label{sec:bg-compgi}
\begin{figure}
    \centering
    \includegraphics[width=8.5cm]{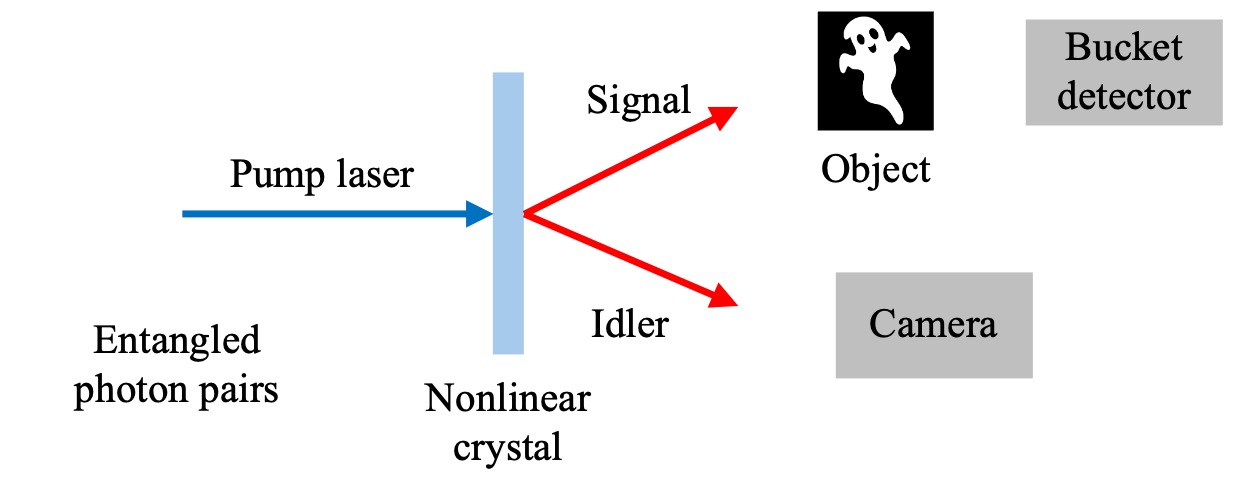} \\
    (A) \\
    \includegraphics[width=8cm]{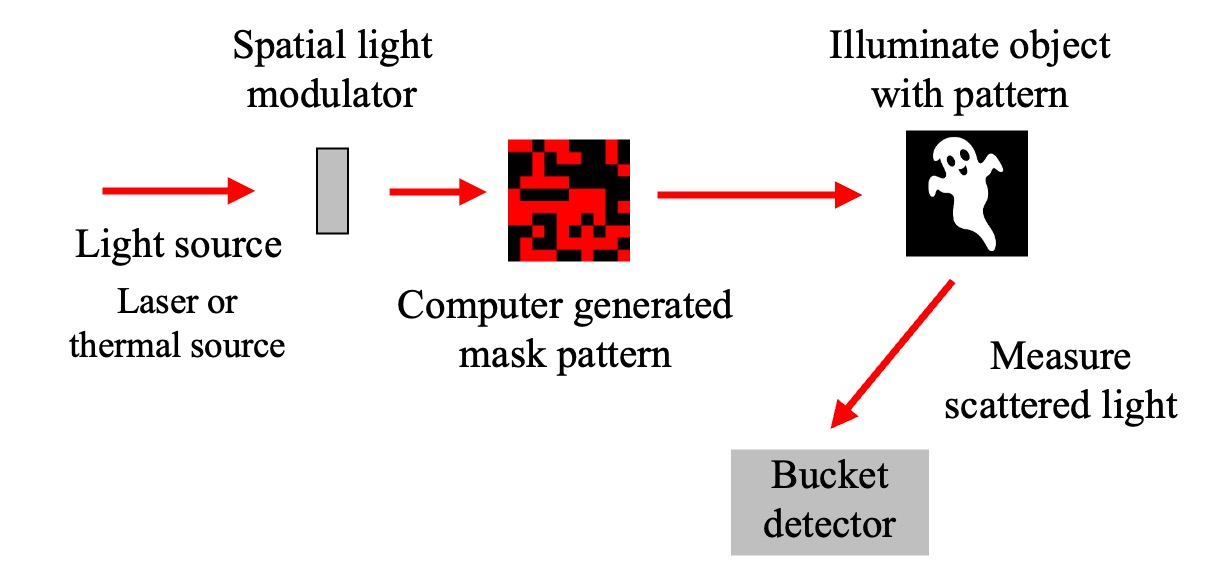} \\
    (B) \\
    \caption{(A) Ghost imaging experimental set-up using entangled photon pairs, (B) Computational ghost imaging.}
    \label{fig:cgi}
\end{figure}

A ghost imaging experiment is implemented such that one light field interacts with the object of interest and is recorded using a bucket detector (or single-pixel detector), while a separate but correlated light field falls onto the pixelated image detector \cite{padgett2017introduction}. In quantum GI, the light field consists of entangled photon pairs (see Fig. \ref{fig:cgi}A), while in classical GI structured or patterned illumination is used with a beam splitter. An image of the object can then be formed using the spatial correlation between these measurements, and neither part of these measurements alone can produce the image. Computational ghost imaging works in a similar way to classical ghost imaging, except that instead of two optical beams, the spatial correlations occur between one optical beam and a pattern stored in a computer's memory \cite{padgett2017introduction} (see Fig. \ref{fig:cgi}B). In computational GI, since one light pattern is known, we do not need the image detector (or the camera) anymore and the only detector needed is the bucket detector to record the light transmitted by the object. The image can be recovered by summing over these patterns weighted by the corresponding bucket values. Details of the recovery method are presented in Sec. \ref{sec:GIreconvery}.

\subsection{Deep Learning in Ghost Imaging}
\begin{figure}
    \centering
    \includegraphics[width=7.7cm]{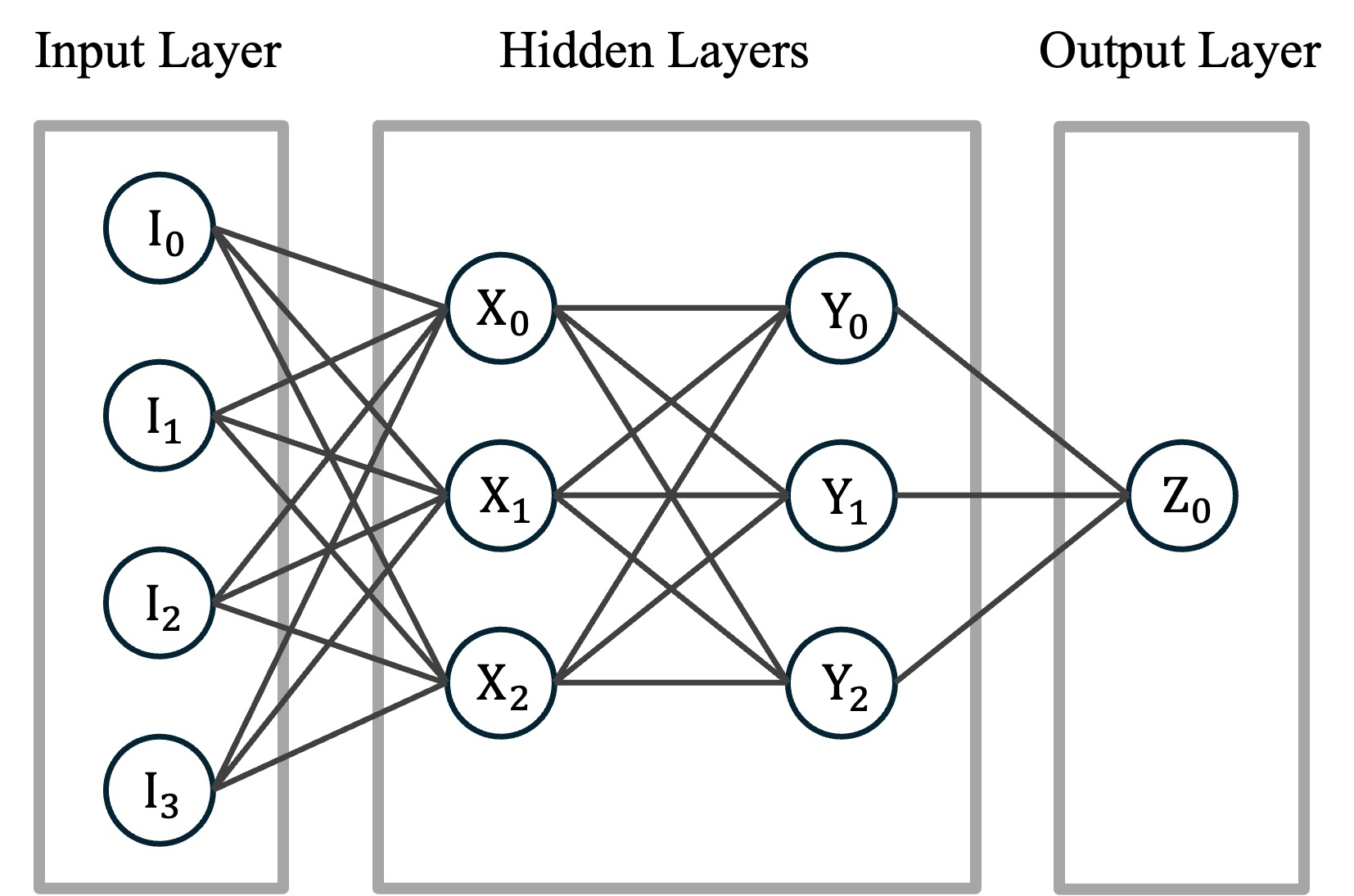}\\
    (A)\\
    \includegraphics[width=8cm]{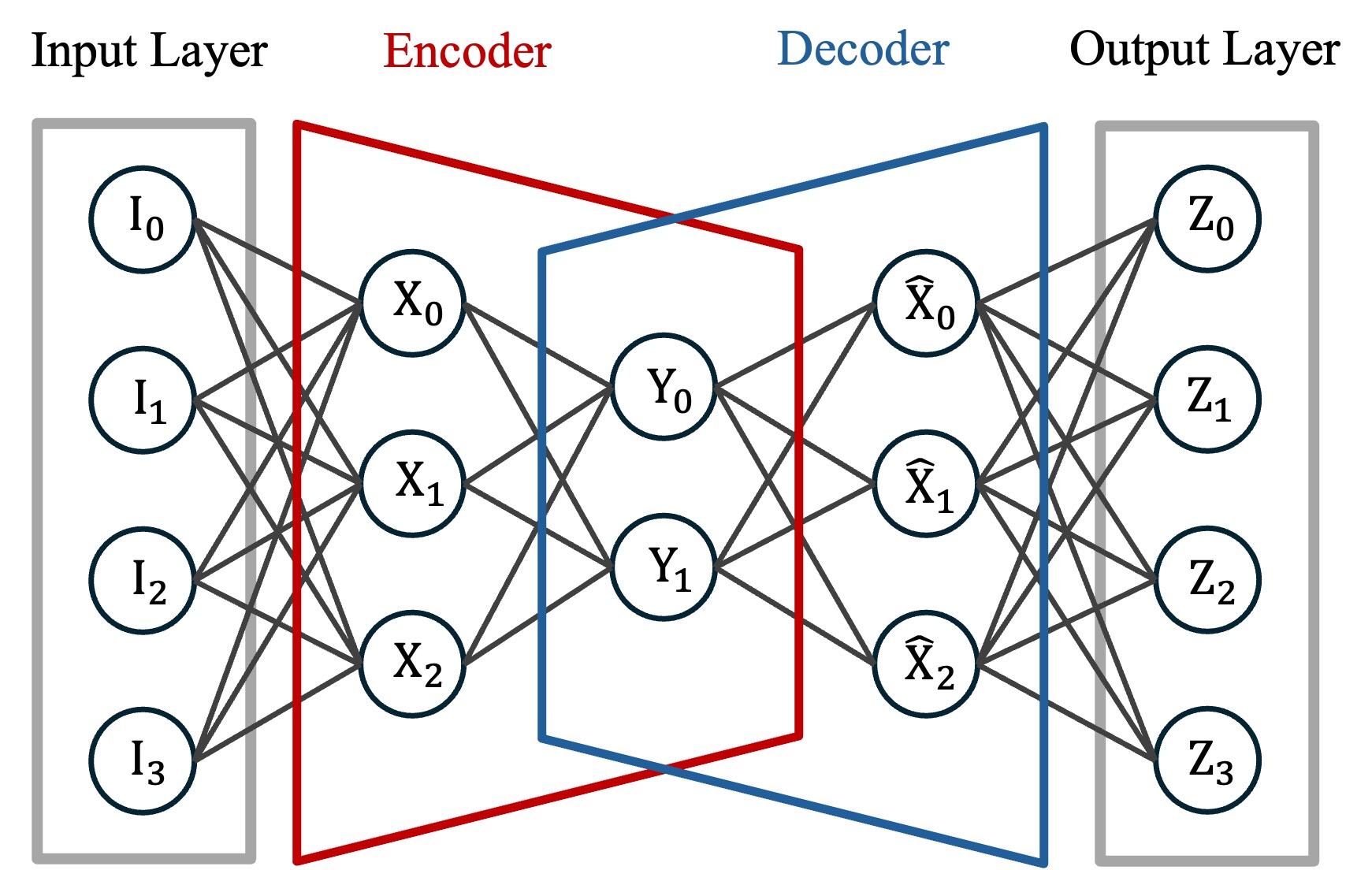}\\
    (B)\\
    \caption{(A) Structure of a simple neural network with two hidden layers (classifiers), (B) Structure of an autoencoder.}
    \label{fig:cf&autoencoder}
\end{figure}

The development of deep learning is inspired by the behavior of the human brain. Figure \ref{fig:cf&autoencoder}A shows the structure of a simple neural network with two hidden layers. Similar to its biological counterpart, an artificial neuron receives a signal, processes it, and sends signals to other neurons connected to it \cite{elgendy2020deep}. Data is ingested into the deep learning model at the input layer, after which it passes through hidden layers where each element's weight and bias are adjusted, and finally to the output layer, where the final prediction is made.
This structure (Fig. \ref{fig:cf&autoencoder}A) is often used as a classifier, in which the classification or pattern recognition is made in the output layer. Figure \ref{fig:cf&autoencoder}B shows the structure of an auto-encoder that aims to match the input provided. It extracts the important features at the encoder part and reconstructs the target at the decoder part. The ResNet \cite{he2016deep}, eHoloNet \cite{wang2018eholonet}, and U-Net \cite{ronneberger2015u,shimobaba2018computational} explored in this work are types of autoencoders based on convolutional neural networks (CNN).

In ghost imaging, if we have prior knowledge of an object of interest, for example, a dataset of images in the same category of the object, we can construct and train a neural network to recover or improve a ghost image. With time, the network will be able to improve image reconstruction accuracy based on the training dataset. This trained network can then be used to recover or recognise the image of our object.

\subsection{Noise Model}
\label{sec:noisemodel}
In x-ray measurements, noise models are very important for statistical image reconstruction, especially for low-dose scans, performing realistic simulations, and assembling lower-dose measurements \cite{bae2001ct,whiting2006properties,nuyts2013modelling}. Numerous sources of noise can affect measurement statistics, but the statistical phenomena of x-ray measurements are so complex that some log-likelihood models are not practical \cite{siewerdsen1997empirical}. On the basis of the statistics in practice, some simple approximations are adequate, such as the Poisson Gaussian model \cite{nuyts2013modelling}. For that reason, we mainly consider two types of noise in x-ray ghost imaging:\\
(1) Photon shot noise

Photon shot noise is due to the random arrival of photons, which may be associated with the quantum nature of light \cite{schoelkopf1998observation}. The number of transmitted photons is hypothesised to follow a Poisson distribution for a given tube current \cite{nuyts2013modelling}. Assume a random number of photons transmitted, denoted by $N$, with a mean value $\bar{N}$ proportional to that current. The probability density distribution follows
\begin{align}
    &N\sim Poisson(\bar{N}),i.e.,\nonumber\\
    &P(N=k)=\frac{\bar{N}^{k}e^{-\bar{N}}}{k!},k=0,1,2,...
    \label{eq:poisson}
\end{align}
and the variance of $N$ equals to its mean $\bar{N}$.\\
(2)	Electronic noise

Electronic noise is caused by leakage current fluctuations in the photosensor and noise in the input of preamplifiers \cite{knoll2000radiation}. A reasonable model for this is a Gaussian distribution; assume we have the equivalent of $\sigma$ photons standard deviation, with the mean centered at the origin since it can be measured and subtracted, then the probability density distribution is
\begin{align}
    &Gaussian(0,\sigma),i.e.,P(x)=\frac{1}{\sqrt{2\pi\sigma^{2}}}e^{-^{\frac{x^{2}}{2\sigma^{2}}}},
    \label{eq:gaussian}
\end{align}
for a random variable $x$.

\subsection{Principal Component Analysis}
\label{sec:PCA}
Principal component analysis (PCA) is a multivariate statistical technique that aims at extracting the most important information from a data set. It represents the data as a set of new orthogonal variables called principal components \cite{abdi2010principal}. In terms of mathematics, PCA is based on eigen-decompositions of positive semi-definite matrices and singular value decomposition (SVD) of rectangular matrices \cite{abdi2010principal}. Let an $n\times m$ data matrix be $X$, then according to SVD we can express $X$ in the form of
\begin{align}
    X=U\Sigma V^{T},
\end{align}
where $U$ is the $n \times l$ matrix of left singular vectors, $l$ is the rank of matrix $X$ with $l \leq \mathrm{min}(n,m)$, $\Sigma$ is the diagonal matrix of singular values, and $V$ is the $m\times l$ matrix of right singular vectors.

In general, the singular vectors are ordered by sorting the singular values from the highest to the lowest, with the highest singular value at the top left of the matrix $\Sigma$ \cite{dash2014principal}. The right singular vectors $V$ represent the principal directions, hence the principal components are given by $XV=U\Sigma$.

\section{Classical Ghost Imaging}
\label{sec:CGI}
In this section, we discuss ghost image recovery using the most commonly used algorithm, where deep learning is not yet involved. We examine the recovered images with different sets of illumination patterns and look for the optimal patterns that allow us to reconstruct the object image with the fewest measurements possible.

\subsection{Method}
\label{sec:CGImethod}
\subsubsection{Classical Ghost Imaging Recovery}
\label{sec:GIreconvery}
In this paper, we focus on two-dimensional ghost imaging. Let the transmission image of the object of interest be $T(x,y)$, where $x$ and $y$ are transverse coordinates at the detector plane. Then, given a set of illumination patterns (or masks) $I_n (x,y)$, where $n=1,2,…,B$ and $B$ denotes the total number of masks used, the one-dimensional set of bucket signals $b_n$ can be modeled as
\begin{align}
    b_n=\sum_{x}\sum_{y}I_n(x,y)T(x,y).
    \label{eq:bucketsignals}
\end{align}

Using the correlation between these sets, we can reconstruct the transmission image according to the principle of classical ghost imaging. The most commonly used GI recovery equation \cite{katz2009compressive,bromberg2009ghost} is given by
\begin{align}
    \hat{T}(x,y)_{GI}=\sum_{n}I_n(x,y)(b_n-\left \langle b \right \rangle),
    \label{eq:transmission}
\end{align}
where $\left \langle b \right \rangle$ denotes the mean value of the set of bucket signals. This is the adjoint operation of the mean corrected imaging process in Eq. (\ref{eq:bucketsignals}).

\subsubsection{Dataset for Simulations}
We used images from the handwritten number set found in Ref. \cite{lecun1998gradient}, with an image size of $28\times 28$ pixels. The handwritten number set comprises two datasets, a training dataset that contains 60,000 images and a testing dataset that contains 20,000 images. The (noise-free) bucket signals were simulated for both the training and testing dataset using Eq. (\ref{eq:bucketsignals}), given the corresponding set of masks. To design a unique set of masks for the number set using the PCA method described in Sec. \ref{sec:PCA}, we used the first 20,000 images from the training dataset for simplicity. 

\subsubsection{Illumination Patterns}
\begin{figure}
    \centering
    \includegraphics[width=\linewidth]{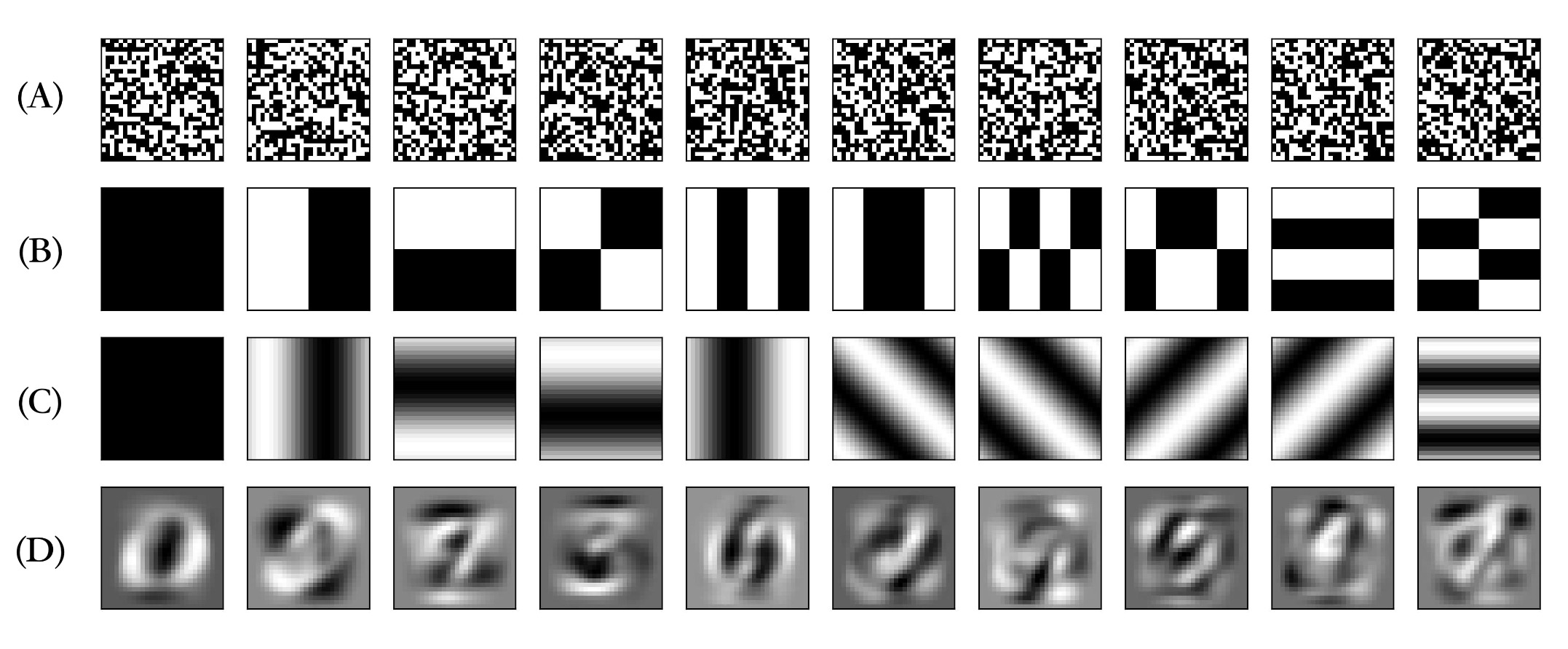}
    \caption{(A) The first ten random binary masks, (B) The first ten lowest-frequency masks of the Hadamard set, (C) The first ten lowest-frequency masks of the Hartley set, (D) The first ten PCA masks, constructed using the first ten most-important components of the dataset.}
    \label{fig:masks}
\end{figure}

For GI, the choice of illumination patterns is crucial. For example, suppose our object only has features near the center of the masks, but our set of masks only contains a series of pinholes around the edges, we might completely miss the feature we are interested in. Here we mainly explored four different types of masks (see Fig. \ref{fig:masks}):\\
(1) Random binary masks, in which half of the pixels are white and the other half are black (Fig. \ref{fig:masks}A). Because of its randomness, this is ideal for compressed sensing.\\
(2) Hadamard masks (Fig. \ref{fig:masks}B). This set of binary masks is constructed to be orthogonal, ordered from low frequencies to high frequencies. \\
(3) Hartley masks (Fig. \ref{fig:masks}C). This set of orthogonal masks is constructed based on sinusoidal functions, which are similar to Fourier basis functions. The set is ordered from low frequencies to high frequencies.\\
(4) PCA masks (Fig. \ref{fig:masks}D). By using principal component analysis to extract the important information of the dataset, this is an orthogonal set of masks designed specifically for the given dataset.

The average transmissibility for these four mask types is about 50\%: the binary masks (random and Hadamard) have an average transmissibility of exactly 50\%, the Hartley mask has slightly higher transmissibility (50\% to 55\%), and that of the PCA mask is slightly lower (45\% to 50\%). We first note that given a set of $B$ masks, lower $B$ would lead to less dose. We require the optimal set of masks that can reconstruct the object image with the minimal number of masks $B$, or the minimal sampling rate $\beta=B/(28\times28)$. We reconstructed the image using the traditional GI recovery method (Eq. (\ref{eq:transmission})), where no deep learning technique is used in this stage.

\subsection{Results and Discussion}
\begin{figure*}
    \centering
    \includegraphics[width=\linewidth]{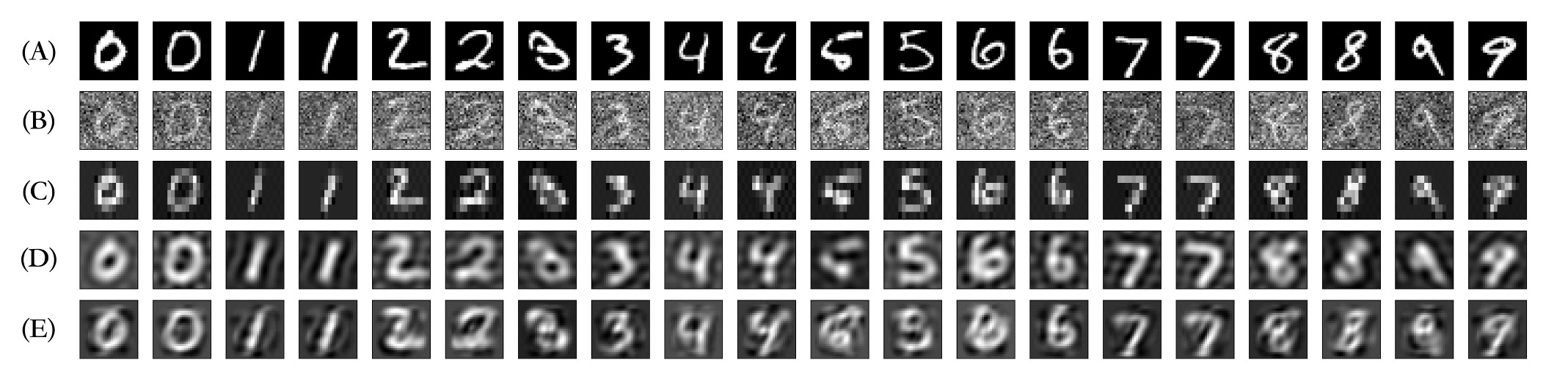}
    \caption{(A) Original images, (B) Classical GI reconstructed images using 294 random binary masks, (C) using 98 Hadamard masks, (D) using 49 Hartley masks, (E) using 20 PCA masks.}
    \label{fig:CGIreconstructions}
\end{figure*}

Figure \ref{fig:CGIreconstructions} shows the reconstructed images using the classical GI recovery method with a given set of masks of length $B$, compared with the original images (Fig. \ref{fig:CGIreconstructions}A). We explored various combinations of these three sets of masks with different $B$ values and found the approximate minimal $B$ value for each group of masks that still produces reasonable results. For the random binary set $B$ = 294, which was at 37.5\% sampling rate (Fig. \ref{fig:CGIreconstructions}B), for the Hadamard set $B$ = 98, $\beta$ = 12.5\% (Fig. \ref{fig:CGIreconstructions}C), for the Hartley set $B$ = 49, $\beta$ = 6.25\% (Fig. \ref{fig:CGIreconstructions}D), and for PCA masks $B$ = 20, $\beta$ = 2.55\% (Fig. \ref{fig:CGIreconstructions}E). The masks were taken in order, i.e., the first $B$ masks were used for each set, except for the random ones, which were selected randomly. The number of masks is chosen to produce approximately the same classifiability in the reconstructed images.

PCA masks performed the best among these types of masks, but if we lowered the sampling rate further, the numbers became unclassifiable, as did the results for other masks. We also noticed that although the set of random masks should be useful from the compressed sensing point of view \cite{candes2008introduction}, they produce quite noisy images, even at a the sampling rate of 37.5\%, and the number of masks cannot be significantly reduced even with compressed sensing applied.

Note that the last two orthogonal sets of masks were at a high bit depth (32 bits per pixel or higher). In practice, achieving such a high bit depth may not be feasible, which could negatively impact the reconstruction results to some extent. Practically speaking, Hadamard masks are likely favoured for their constructibility, although theoretically, we can see that we would need at least twice the number of masks compared to the Hartley and PCA cases to achieve classifiable results.

From here on, we focus on the two best-performing orthogonal sets of masks, namely Hartley and PCA, which have greater potential for dose reduction.

\section{Deep Learning in Ghost Imaging}
\label{sec:DLGI}
In the previous section, we showed how choosing the optimal illumination pattern could improve the reconstructed images' quality or reduce the number of patterns required, but the effect was still limited, especially when the sampling rate was low. In this section, we investigate how deep learning techniques can improve the quality of images subject to low sampling rates. We propose a neural network that reconstructs images directly from the acquired bucket signals, then compare the results with those reconstructed using the adjoint method in Sec. \ref{sec:CGI}.

\subsection{Network Structure}
\begin{figure*}
    \centering
    \includegraphics[width=\linewidth]{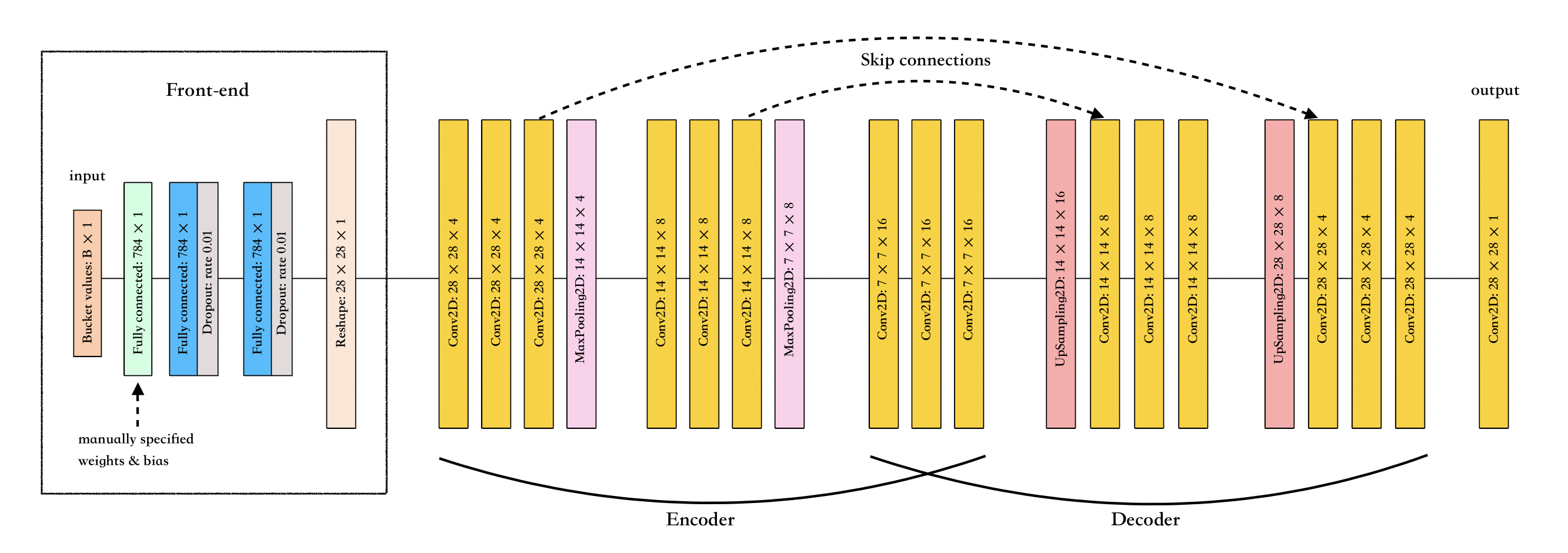}
    \caption{Our end-to-end neural network structure, which can reconstruct object images directly from the measured bucket signals.}
    \label{fig:network}
\end{figure*}

Here we propose an end-to-end neural network to compensate for the missing information due to low sampling rates. Its structure is shown in Fig. \ref{fig:network}. The network takes a one-dimensional normalised bucket value list of length $B$ as input and outputs the predicted two-dimensional object image. We call the network for receiving the input to reshape it into 2D image information the “Front-end”, with the CNN autoencoder used to denoise the images being called the “Back-end”. In the Front-end we mainly used fully connected layers to map the bucket value list to the full image size. From the adjoint GI recovery algorithm (Eq. (\ref{eq:transmission})) we know how each bucket value contributes differently, i.e., the weight of which varies. Hence, instead of giving randomly generated weights and biases, we manually set these values to give our network a better starting point. We set the weights of the first fully connected layer according to Eq. (\ref{eq:transmission}). They were normalised by the GI adjoint scale $\gamma^{-1}$ derived in Ref. \cite{kingston2021inherent}, where $\gamma=B\sigma_I^{2}$; here $\sigma_I$ denotes the variance of the set of illumination masks. Note that the derivation for $\gamma$ relies on the random-pattern properties, however, this is still approximately correct for other types of mask in practice. Accordingly, we adjusted the bias so that the final images would fall within a range of 0 to 1, as preferred for DL networks. Two more fully connected layers were added for optimisation, each followed by a dropout layer with a rate of 0.01 to prevent overfitting. The data were then reshaped into a two-dimensional format before entering the Back-end. As such, the Front-end’s main function is to reshape the 1D bucket values measured into 2D images, which can still produce object images from bucket signals even without the Back-end. However, the image quality will not be as good. Inspired by U-Net \cite{ronneberger2015u,shimobaba2018computational}, our Back-end employed a convolutional neural network (CNN) to improve the image quality. The Back-end was composed only of convolution, max-pooling (down sampling process), and up-sampling layers. Each convolution layer used a Rectified Linear Unit (ReLU) as its activation function, which removes negative data. Each convolution layer had a convolution kernel size of $3\times 3$ pixels. The down-sampling rate was $2\times 2$ pixels in both max-pooling layers, and accordingly the same rate in the up-sampling layers. To avoid resolution loss in max-pooling layers, skip connections were added to the network so that high-resolution details can be preserved for predictions. The final convolution layer yielded our predicted object image. The network was implemented using Keras in the TensorFlow platform \cite{abadi2016tensorflow}. 

\subsection{Network Training}
\label{ssec:noisefreeTrain}
As previously mentioned, the training of the network is where the weights and bias factors of each neuron in the network are adjusted and optimised. In this section, we used the first 20,000 images in the training dataset for training, and the first 4,000 images in the testing dataset for testing. The network was trained by minimising the mean squared error (MSE) between its output and the corresponding original images for each bucket value. We used the embedded Adam optimiser \cite{jais2019adam} in Keras for optimisation with a learning rate of 0.0001. We set the batch size to 128 and set training epochs to 300. This set of hyperparameters was approximately optimal among a wide range of sets of masks used, without causing significant overfitting problems in our cases.

\subsection{Results and Discussion}
\begin{figure*}
    \centering
    \includegraphics[width=\linewidth]{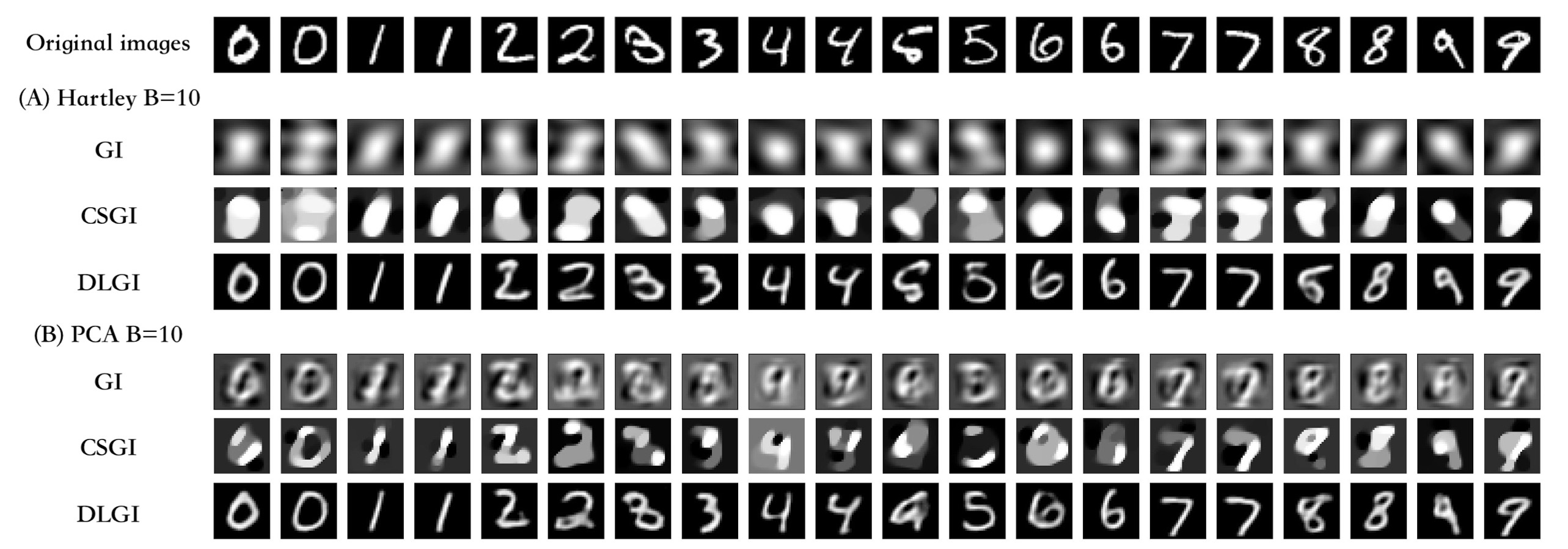}
    \caption{Reconstructed images using the proposed network (DLGI) compared with images recovered using the traditional method (GI) and the compressed sensing method (CSGI). Top row: original images. (A) case with 10 Hartley masks, (B) case with 10 PCA masks.}
    \label{fig:DLGIreconstructions}
\end{figure*}

Figure \ref{fig:DLGIreconstructions} shows the reconstructed images obtained using our network (labelled as DLGI), the classical GI method (labelled as GI), and the NESTA algorithm \cite{becker2011nesta}, a compressed sensing based method focusing on minimising the total variation (labelled as CSGI), for two cases. With Hartley and PCA masks, our network was capable of reconstructing handwritten digits accurately. Even when $B$ was reduced to 10 ($\beta$ = 1.28\%), most images still resembled the original images. For Hartley masks, we ended up choosing the first ten lowest-frequency masks, as shown in Fig. \ref{fig:masks}B. We also examined random selection spanning all frequencies, as well as selections from each frequency range (i.e., one randomly chosen from the first ten, one from the next ten, etc.). However, the first ten lowest-frequency masks yielded the best results in this case.

The results demonstrate that deep learning is a powerful tool for reconstructing ghost images. We can see that especially in Fig. \ref{fig:DLGIreconstructions}A, while the traditional GI recovery method and the compressed sensing method gave barely any information, the quality of the images reconstructed using our network was still promising. For the Back-end, we also exploited ResNet \cite{he2016deep} with 50 layers, and a similar structure to eHoloNet \cite{wang2018eholonet}, the U-Net (used in our network) produced overall the best results for the handwritten number set. For other datasets, especially with more complex images, other structures might outperform U-Net, and we would probably need to expand our network structure to have more convolutional layers to accommodate complex datasets.

\section{Noise Simulation and Dose Reduction}
So far, the simulations have been conducted without noise, but as mentioned in Sec. \ref{sec:noisemodel}, in experimental x-ray ghost imaging two types of noise need to be considered. The noise was taken into account in this section, with the same network as in Sec. \ref{sec:DLGI} being applied to simulated bucket signals with noise added. The resulting images were then compared to those formed by direct imaging (conventional camera) subjected to the same amount of total dose, to see if dose reduction is possible using DLGI.

\subsection{Method}
We let the standard deviation of the Gaussian noise be the equivalent of $\sigma$ photons and let the incident flux be $p$ photons per pixel per mask. Using Eq. (\ref{eq:poisson}) and Eq. (\ref{eq:gaussian}), we then simulated the bucket signals list with noise added as
\begin{align}
    b_n^{noise}=\frac{Poisson(p\cdot b_n)+Gaussian(0,\sigma)}{p},
\end{align}
where $b_n$ is the noise-free bucket signal.

In direct imaging, which is the traditional method of x-ray imaging, every pixel is illuminated at the same time. If a given set of masks with length $B$ has an average transmissivity of $\alpha$, then the photon flux per pixel that gives the same experiment dose on the object as GI would be $\alpha\cdot p\cdot B$. We then modified each original pixel value $x$ to be
\begin{align}
    x^{noise}=\frac{Poisson(\alpha\cdot p\cdot B\cdot x)+Gaussian(0,\sigma)}{\alpha\cdot p\cdot B}.
\end{align}

Using a low dose, $p$ = 5, and a fixed Gaussian standard deviation, $\sigma$, we examined how our DLGI image quality compared to that of direct imaging subject to the same total amount of dose. We focused on using Hartley masks and PCA masks with $B$ = 10, whereby the average transmissibility for these two sets were 0.535 and 0.486, so the same amount of total dose in direct imaging would be approximately 4194 photons and 3810 photons respectively. 

\subsection{Datasets}
In this section, we used two sets to test the performance of our neural network: the handwritten number set as presented in the previous sections, and a more complicated fashion set \cite{xiao2017fashion}. We chose simple images, i.e., the numerals, to explore classifiability (can we tell which numeral it is?) and more diverse images with detailed information, i.e., the fashion dataset, to explore reconstructed image fidelity (measured through mean squared error). Both datasets consist of 60,000 training images and 20,000 testing images, with grayscale images of size 28 
$\times$ 28. Here, we used the full datasets for our network training and testing. The network training details were the same as presented in Sec. \ref{ssec:noisefreeTrain}, with training conducted for the noisy bucket values and the corresponding original images.

\subsection{Results and Discussion}
\label{sec:5C}
Before commencing the discussion, we note that all comparisons made between direct imaging and our DLGI results were based on the assumption that both approaches have the same detector quantum efficiency, with DLGI being at an inherent disadvantage. More details on this are provided at the end of this section.

\begin{figure*}
    \centering
    \includegraphics[width=19cm]{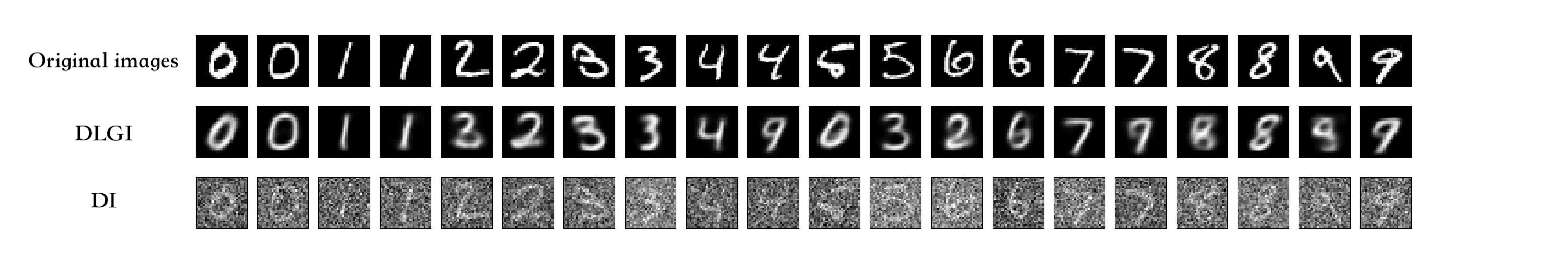}
    \includegraphics[width=19cm]{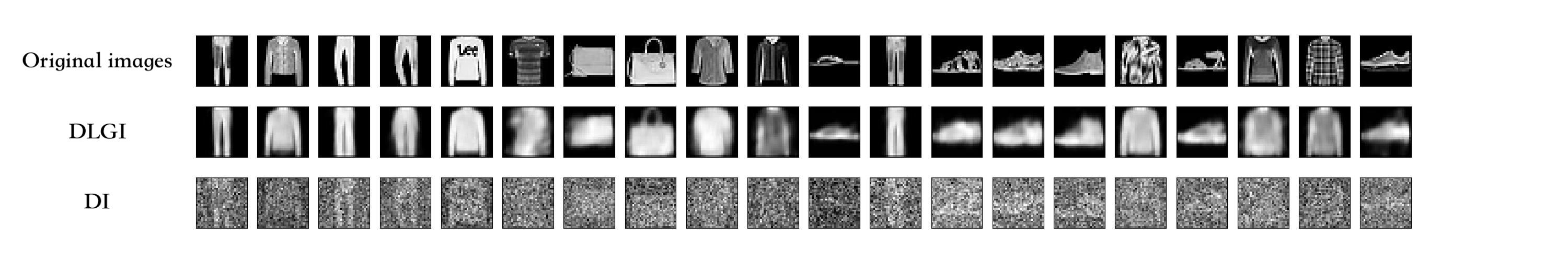}
    \caption{Reconstructed number and fashion GI images using our network with 10 PCA masks (DLGI), compared with the images captured using direct imaging (DI) subjected to the same total dose of $\sim 3810$ photons. Gaussian noise standard deviation $\sigma$ = 15.}
    \label{fig:10pcanoise}
\end{figure*}

Figure \ref{fig:10pcanoise} illustrates the number and fashion GI images recovered using our network (DLGI) compared to direct imaging (DI) with the same dose, along with the original dataset. Note that we omitted results with Hartley masks, as they perform similarly to PCA masks. In contrast to the DI results, which are extremely noisy, the DLGI results seem to have a good chance of reconstructing most numbers due to the design of the network. For the handwritten number case, we noticed that although the image could still be recognised as a number instead of other objects, some of the numbers were incorrectly recognised. For the fashion dataset case, one limitation of our neural network is that it did not preserve details well for complicated datasets; we can see that the letters and stripes on the clothes are missing and blurred. The issue persisted even with an increased dose, i.e., an increased number of masks. Therefore, one potential way to further improve image quality would be to add more layers to the network when dealing with more complex images.

\begin{figure*}
    \centering
    \includegraphics[width=19.0cm]{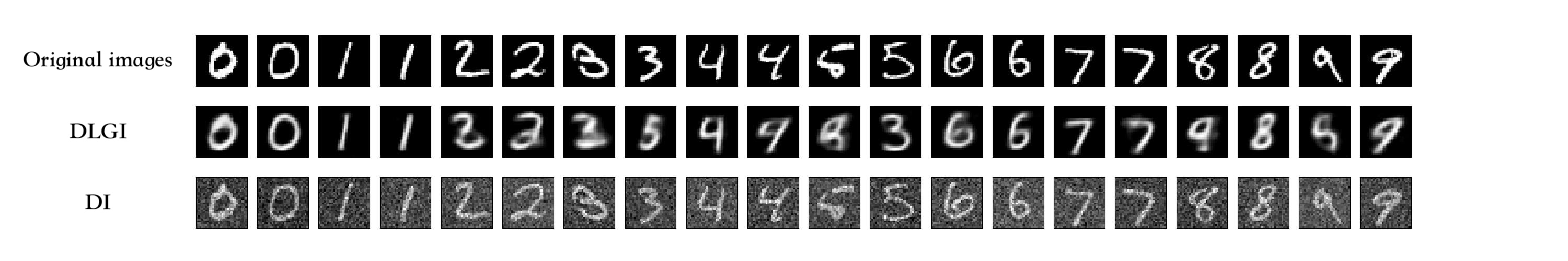}
    \caption{Reconstructed number GI images using our network with 10 PCA masks (DLGI), compared with the images captured using direct imaging (DI) subjected to the same total dose of $\sim 3810$ photons. Gaussian noise standard deviation $\sigma$ = 5.}
    \label{fig:10hartley_sig5}
\end{figure*}

Since we assumed that DI was dominated by Gaussian noise, if the standard deviation in the Gaussian noise could be reduced, DI results would be better and easier to distinguish. A potential advantage of DLGI is that its result will not be dramatically affected by the amount of Gaussian noise in experiments, i.e., the reconstructed image quality will be around the same level whether the $\sigma$ value is small or large relative to the Poisson noise (see Fig. \ref{fig:10hartley_sig5} for hand-written number set with $\sigma$ = 5).

\begin{figure*}
    \centering
    \includegraphics[width=19.0cm]{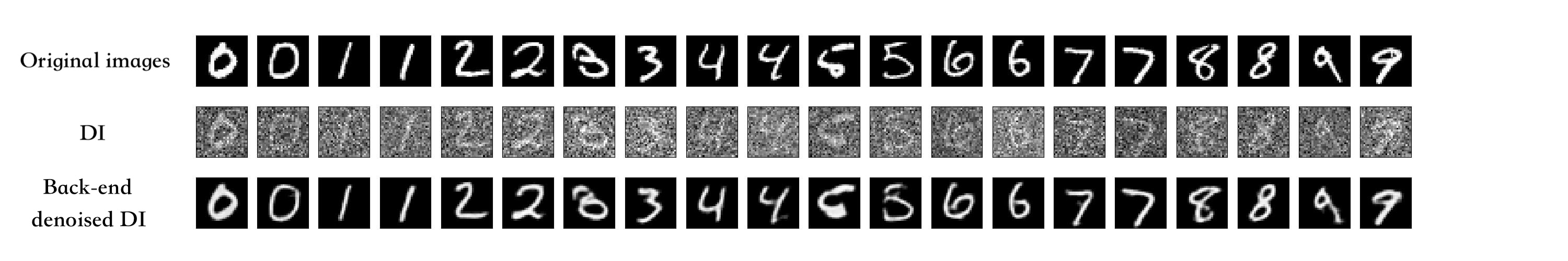}
    \includegraphics[width=19.0cm]{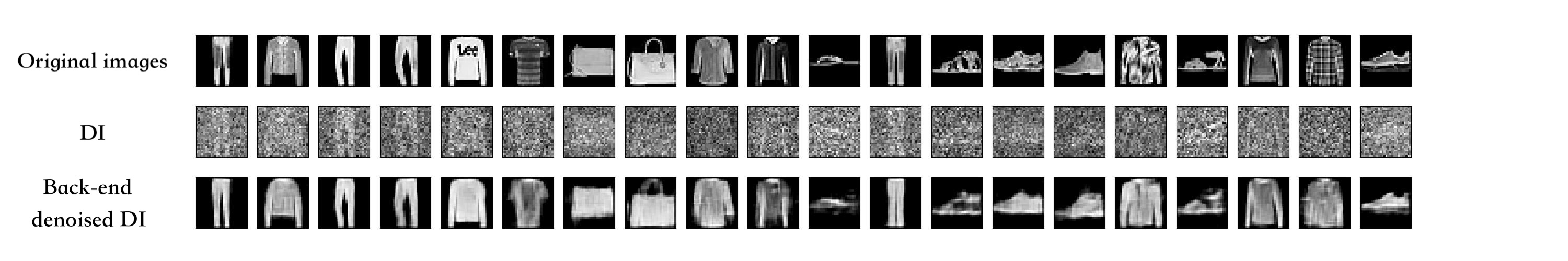}
    \caption{Back-end denoised direct images with a total dose of $\sim 3810$ photons, Gaussian noise $\sigma$ = 15.}
    \label{fig:denoisednumberfashion}
\end{figure*}

Note that we can always denoise DI results using deep learning or other techniques if we have prior knowledge about the object of interest as well. We performed the denoising using the back-end presented in Sec. \ref{sec:DLGI} on the DI results shown in Fig. \ref{fig:10pcanoise}. The denoised results are shown in Fig. \ref{fig:denoisednumberfashion}. The details not preserved well in the denoised fashion dataset emphasise that the current network structure needs proper modification to accommodate finer feature images.

\begin{figure*}
    \centering
    \begin{minipage}{0.33\linewidth}
        \centering
        \scriptsize{(A)}\\
        \includegraphics[width=\linewidth]{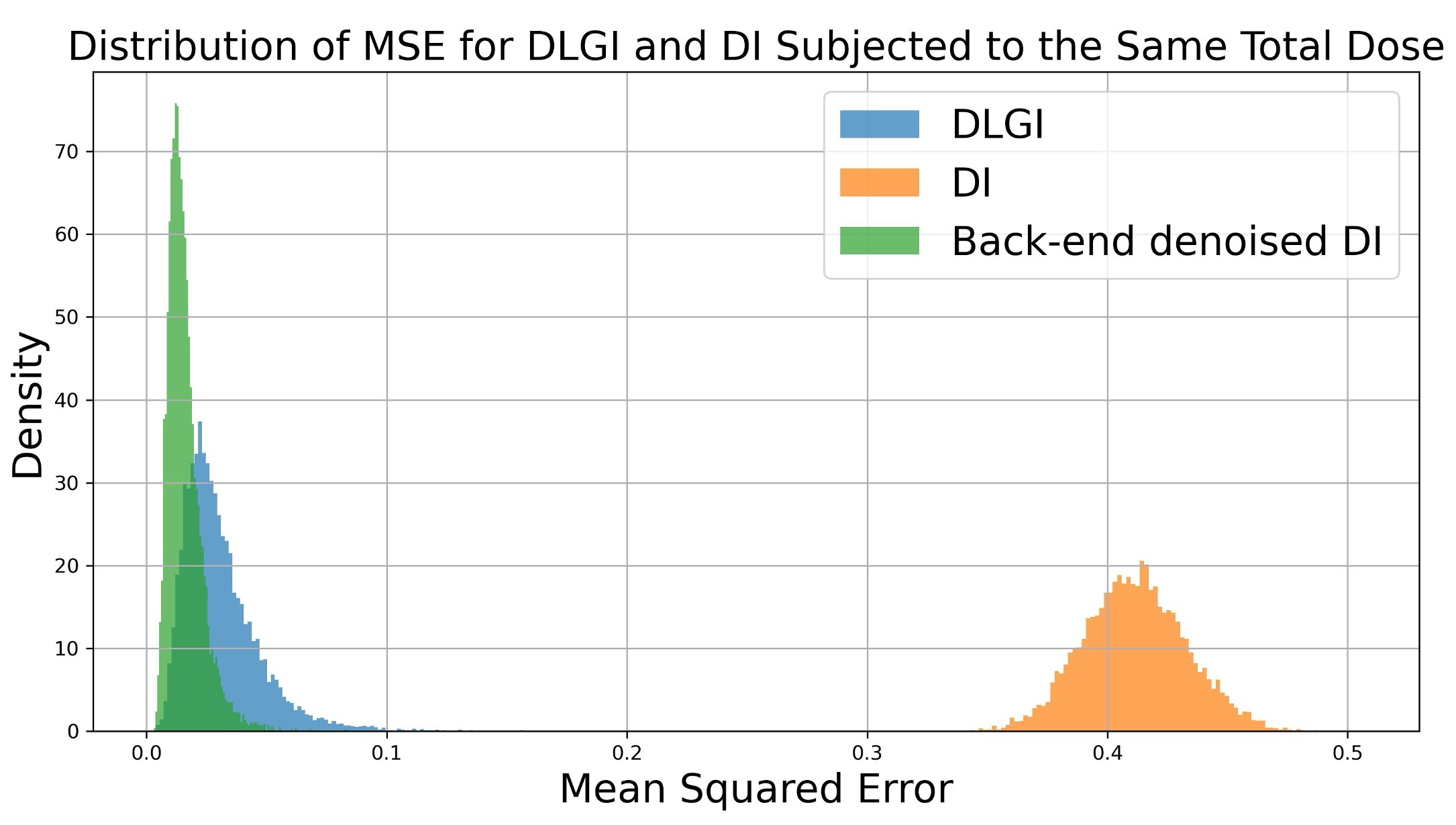}
    \end{minipage}%
    \begin{minipage}{0.33\linewidth}
        \centering
        \scriptsize{(B)}\\
        \includegraphics[width=\linewidth]{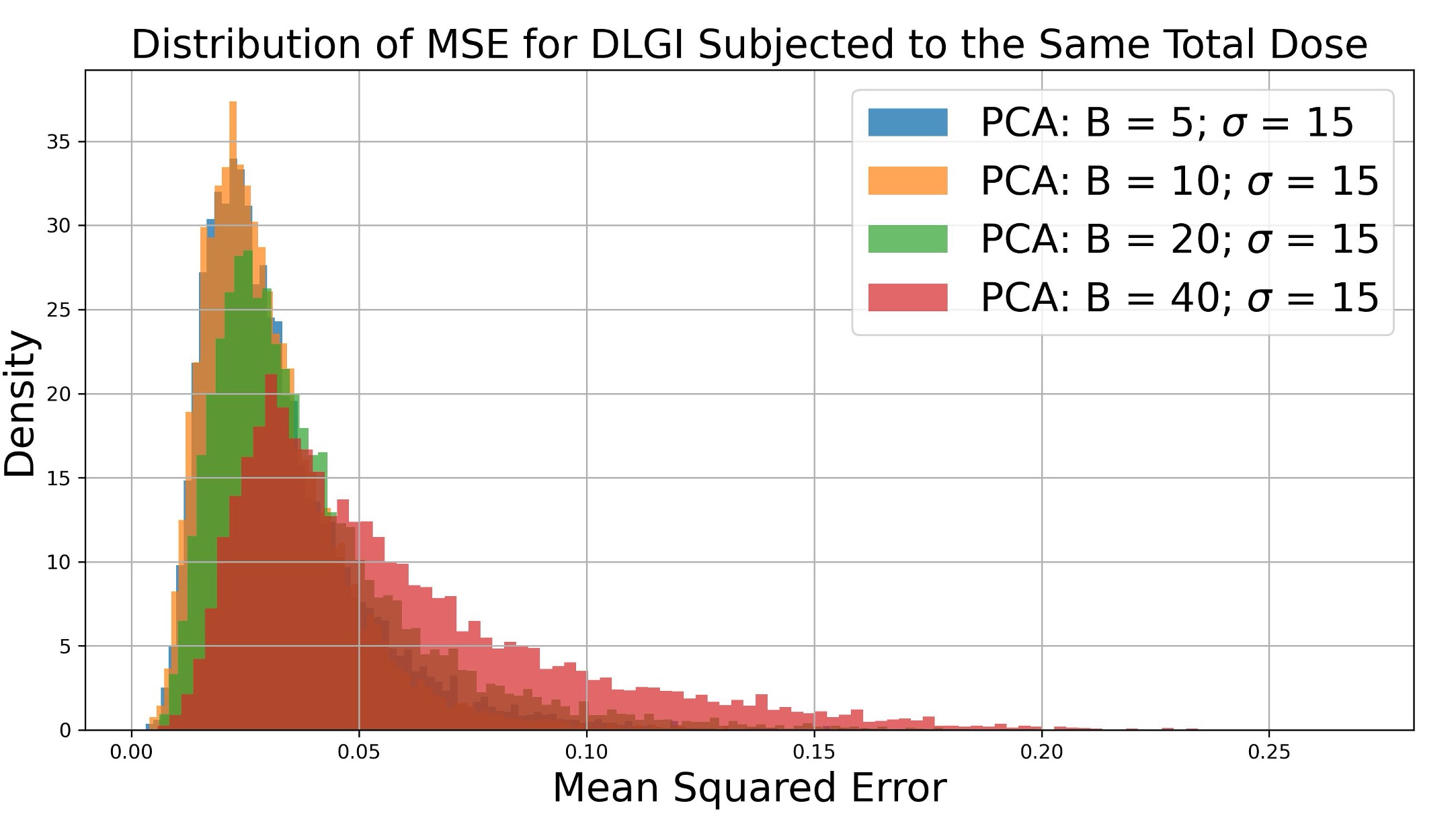}
    \end{minipage}%
    \begin{minipage}{0.33\linewidth}
        \centering
        \scriptsize{(C)}\\
        \includegraphics[width=\linewidth]{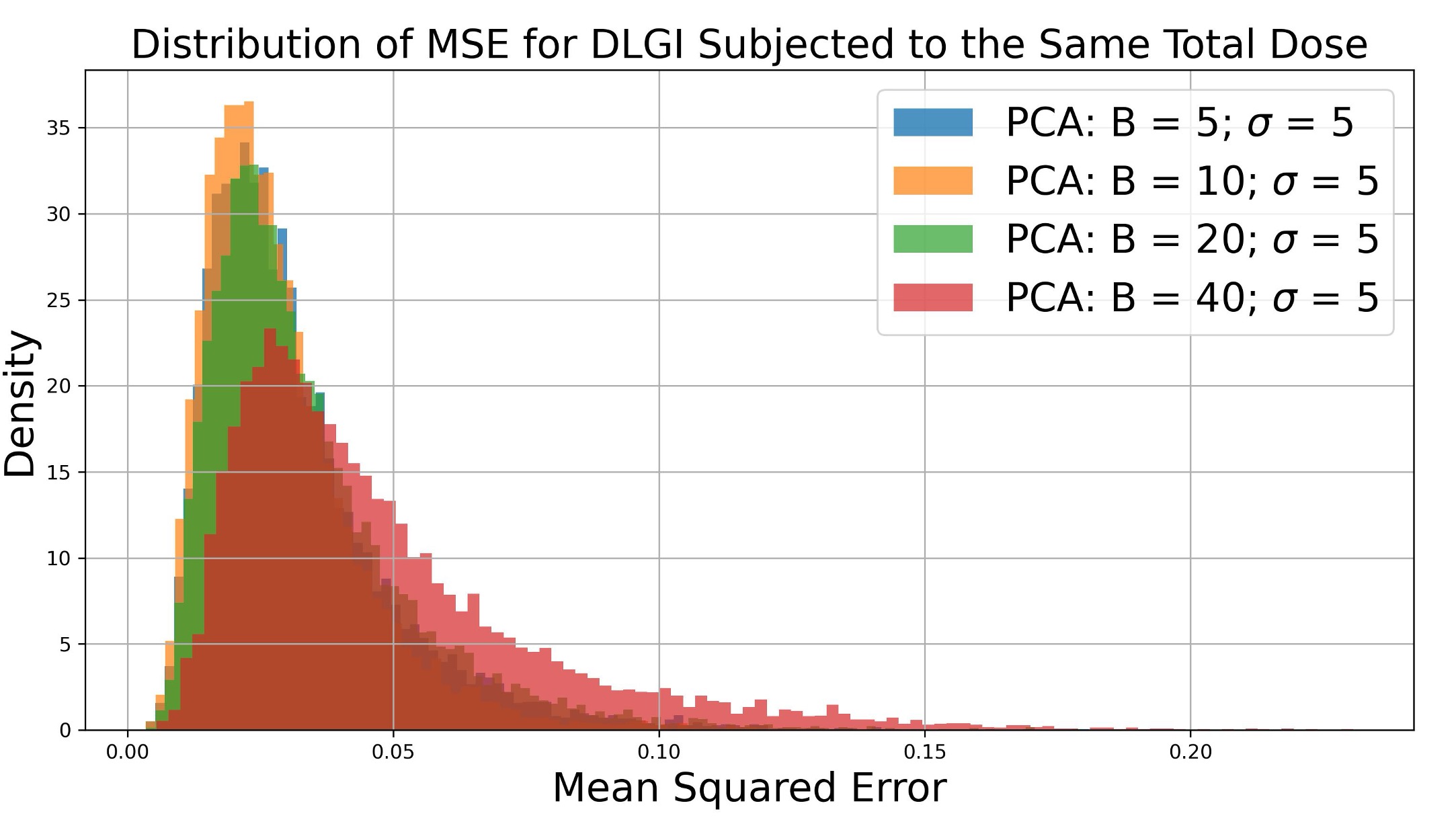}
    \end{minipage}
    \caption{Distribution of mean squared error for the fashion dataset (10,000 test images) subjected to the same total dose of $\sim 3810$ photons. (A) Comparison between DLGI reconstructed images with 10 PCA masks, DI images, and back-end denoised DI images. Gaussian noise $\sigma$ = 15; (B) Comparison between DLGI images with Gaussian noise $\sigma$ = 15, using 5, 10, 20, and 40 PCA masks respectively; (C) Same comparison as (B) but with Gaussian noise $\sigma$ = 5.}
    \label{fig:mse}
\end{figure*}

Figure \ref{fig:mse} shows the distribution of mean squared error (MSE) for the fashion dataset (10,000 test images) subjected to the same total dose of around 3810 photons. From Fig. \ref{fig:mse}A, we can see that DLGI achieved a much smaller MSE compared to DI. It is also worth noting that using the traditional GI recovery method (refer to Sec. \ref{sec:CGI}), the reconstructed images obtained an average MSE of 32.97 with a standard deviation of 21.89, which was significantly improved using deep learning. When given the same prior knowledge, the DI images denoised using the back-end of the network performed the best overall, suggesting that it is very challenging for DLGI to surpass DI with the same trainable network (or equivalent prior knowledge) under extremely low-dose conditions. We also checked the network performance when distributing a total dose of 3810 photons to a different number of PCA masks used. The results are shown in Fig. \ref{fig:mse}B and \ref{fig:mse}C for $\sigma$ = 15 and 5, respectively. Overall, under the same conditions, less Gaussian noise ($\sigma$ = 5) performed similarly but slightly better than higher Gaussian noise ($\sigma$ = 15). For both cases, 10 masks gave the best performance among 5, 10, 20, and 40 masks. Inspecting the trend more carefully, for $\sigma$ = 5, the best performance should be given by 10 masks, since the distributions for 5 and 20 masks roughly overlap with each other; and for $\sigma$ = 15, the best performance should be given by 7 or 8 masks. In practice, determining the optimal number of masks for a given total dose involves a two-step process. First, we can test a few selected mask counts to identify any trend suggesting best performance. Based on these initial results, we then run the deep learning program over all numbers within this range. Although this approach may be time-intensive, it provides valuable insights to guide optimal performance under low-dose conditions.

We did not include a quantitative analysis such as calculating the MSE between the reconstructed images and the original images for the hand-written number case. Although we can do that, this would be less significant here since, for most medically used x-ray images, we expect them to be recognisable to human eyes. For instance, in Fig. \ref{fig:10pcanoise} the first ``2" and the second ``7" were reconstructed to a ``3" and a ``9" respectively. However, if we calculated the MSE of the DLGI reconstructed images with the original images and compared it to that of the DI results, we would still come up with a smaller value for the error. Although the MSE was relatively low, one can still clearly tell that this was an incorrect number. 
In this context, a more suitable metric would be classification accuracy, which quantifies the ability to correctly identify the target numbers. We used a simple classifier network that has a similar structure to the one shown in Fig. \ref{fig:cf&autoencoder}A (refer to details in Appendix). We trained the network with the original training images and tested it with our DLGI/DI results presented in Fig. \ref{fig:10pcanoise}. For the handwritten number set, the accuracy for DLGI was approximately 65\%, while that of DI was around 75\%. For the fashion dataset, both DLGI and DI achieved an accuracy of around 68\%. The image classifiability using DLGI was comparable to conventional direct imaging. Despite 35\% of numbers and 32\% of clothing items being classified incorrectly, these results were still encouraging considering the extremely low dose conditions. Therefore, DLGI shows potential for recovering images under low doses with a low sampling rate, especially when electronic noise is significant.

We have assumed the same detective quantum efficiency (DQE) and electronic noise in this section for the bucket detector used in GI and the pixelated detector used in DI. In practice, \textit{a bucket detector could have a far higher DQE since detector spatial resolution is not required}. There is a blurring component introduced by the use of a scintillator; a thin scintillator will blur less but has a lower detection efficiency \cite{arhatari2023micro}. A compromise is required between efficiency and achievable resolution. For example, consider the following conventional imaging scenarios where dose is important:
\begin{itemize}
    \item A medical chest x-ray of a large adult recorded with 0.139mm pixel size on a flat-panel detector. The x-ray spectrum used could be 125kV accelerating voltage with 2.5mm Al filtering. A flat-panel detector with 4mm carbon-fiber and 0.25mm Al sensor protection followed by the 700$\mu$m CsI scintillator could be employed. In this case, only 74.3\% of the x-ray photons are detected, or 71.7\% of the total energy of incident photons.
    \item A preclinical imaging study of mice recorded with 6.5$\mu$m pixel size on CCD using 100$\mu$m CsI scintillator. An example x-ray spectrum might be 60kVp accelerating voltage with 0.5mm Al filtering. In this scenario, only 56.9\% of the incoming x-ray photons are detected, or 53.6\% of the total energy of the incoming photons.
    \item Scanning a biological sample at a synchrotron with a monochromatic beam of 20keV x-rays using a detector with 1.3$\mu$m pixel (after $5 \times$ optical magnification) size and a 35$\mu$m LuAG (Lu$_3$Al$_5$O$_{12}$) scintillator. Here, only 37.4\% of the incoming x-ray photons are detected.
\end{itemize}
In the case of ghost imaging, pixel size is irrelevant, therefore a large optically-transparent scintillator such as NaI can be used to achieve close to 100\% DQE in all cases. This suggests that DLGI can be a promising method for dose reduction in practice. Taking into account the improved DQE of a bucket detector over a pixelated detector for direct imaging, we anticipate an increase in measured flux in the GI data by 35\%, 75\%, and 160\% in the above example scenarios whilst maintaining the same dose on the object.

\section{Conclusions and Future Work}
We have proposed an end-to-end neural network based on U-net that can reconstruct the ghost images directly from the bucket signals acquired. Using this network we demonstrated the potential for dose reduction in x-ray ghost imaging with deep learning by simulation. We found that orthogonal sets of illumination patterns, including those designed uniquely by the PCA method based on prior knowledge, are useful in traditional GI. In DLGI, the orthogonal sets allowed the images to be reconstructed even when the sampling rate was extremely low ($\beta$ = 1.28\%). Moreover, we found that DLGI can benefit from the fact that the reconstructed image quality will not be significantly affected by the amount of electronic noise. However, it is important to note that this model does not account for out-of-distribution robustness. From a real-world perspective, this means that, for example, we cannot train the network on a set of healthy chest images and then use it to diagnose lung diseases. With prior information about the object of interest, ghost imaging has potential in image reconstruction under low doses with the proper choice of masks and the assistance of deep learning. However, given the same prior knowledge and detector quantum efficiency, it is very challenging for DLGI to outperform DI under extremely low-dose conditions.

A next step would be to adjust the network details to better preserve features for complex datasets such as the fashion set. This can involve modifying the current network structure—e.g., adding more layers—or constructing another structure regarding the features in the target dataset. Moreover, given the same prior knowledge, we observed that denoised DI results were promising, even better than that of DLGI. So, for DLGI to outperform DI we need much more prior knowledge, probably specific to what the object exactly looks like. It is then worth asking, what circumstances would make it worthwhile to use GI rather than DI? 
It would be valuable to test the proposed approach in different real-world scenarios, as mentioned at the end of Sec. \ref{sec:5C}, where we expect GI to obtain an increased flux compared to DI, subject to the same total dose, with the extent of this increase varying depending on the object. Such additional work is needed to properly investigate the efficiency of the methods in the present paper for practical applications.

\begin{acknowledgments}
AMK and DMP thank the Australian Research Council (ARC) for funding through the Discovery Project: DP210101312. AMK and PDL thank the ARC and Industry partners funding the Industrial Transformation and Training  Centre for Multiscale 3D Imaging, Modelling, and Manufacturing: IC180100008. The authors acknowledge helpful discussions with Dr Lindsey Bignell, Dr Wilfred Fullagar, A/Prof Marcus Kitchen, and Dr Andrew Stevenson.
\end{acknowledgments}

\appendix*
\section{Classifier Used in This Work}
\label{app:classifer}
The classifier used to determine the DLGI and DI images' classifiability has the following architecture:

\begin{enumerate}
    \item Input Layer: Converts the 28 $\times$ 28 pixel images into a 1D array (784 dimensions) to feed into the subsequent layers.
    \item Two Hidden Layers: The first hidden Dense layer consists of 784 neurons, and the second consists of 392 neurons, each using the ReLU activation function. Each hidden layer is followed by a Dropout layer with a dropout rate of 0.2 to prevent overfitting.
    \item Output Layer: The Dense layer comprises 10 neurons, each representing a class in the classification task, i.e., digits 0-9 in the handwritten digit dataset and ten different classes in the fashion dataset.
\end{enumerate}

The model is compiled with the Adam optimiser \cite{jais2019adam}, the sparse categorical crossentropy loss function, and the accuracy metric for evaluation. The classifier is trained on the original training dataset with 30 epochs and a batch size of 128.


\bibliography{references}

\end{document}